# A new Bayesian ensemble of trees classifier for identifying multi-class labels in satellite images

Reshu Agarwal, Pritam Ranjan and Hugh Chipman

*Department of Mathematics and Statistics*

*Acadia University, Wolfville, Nova Scotia, Canada*

## Abstract:

Classification of satellite images is a key component of many remote sensing applications. One of the most important products of a raw satellite image is the classified map which labels the image pixels into meaningful classes. Though several parametric and non-parametric classifiers have been developed thus far, accurate labeling of the pixels still remains a challenge. In this paper, we propose a new reliable multiclass-classifier for identifying class labels of a satellite image in remote sensing applications. The proposed multiclass-classifier is a generalization of a binary classifier based on the flexible ensemble of regression trees model called Bayesian Additive Regression Trees (BART). We used three small areas from the LANDSAT 5 TM image, acquired on August 15, 2009 (path/row: 08/29, L1T product, UTM map projection) over Kings County, Nova Scotia, Canada to classify the land-use. Several prediction accuracy and uncertainty measures have been used to compare the reliability of the proposed classifier with the state-of-the-art classifiers in remote sensing.

## INTRODUCTION

Satellite image classification plays a crucial role in numerous remote sensing data applications like land-use planning, land cover change monitoring, forest degradation assessment, hydrological modeling, sustainable development, wildlife habitat modeling, biodiversity conservation, and so on. One of the most important inputs in such an application is the classified map, which is a raw image with pixel-wise class labels. Thus, it becomes essential to achieve the highest possible accuracy of the classified maps. Several classifiers have been developed and implemented worldwide (e.g., Franklin et al., 2002; Pal and Mather, 2003; Gallego, 2004), however, accurate labeling of the pixels still remains a challenge often due to the complexity in study area terrain, sensor characteristics and training data size estimation (Blinn, 2005; Song et al., 2012).

Among various parametric classification methods, the maximum likelihood (ML) classifier has been the most widely used classifier in image processing software (Peddle, 1993). However, the ML classifier is based on a parametric model that assumes normally distributed data which is often violated in complex land-use satellite images (Lu and Weng, 2007). Non-parametric classifiers, which do not make strong assumptions like normality, have gained much popularity. Classifiers based on k-nearest neighbor (k-NN), artificial neural networks (ANN), decision trees and support vector machines (SVM) have shown better performance as compared to ML classifiers for complex landscapes (Zhang and Wang, 2003; Bazi and Melgani, 2006; Li et al., 2010). Of course, the non-parametric methods are not perfect either, and have many shortcomings. Ranking of such classifiers has been an interesting research area in the machine learning literature. For example, Sudha and Bhavani (2012) concluded that SVM is a better classifier than k-NN, and Song et al. (2012) argue that SVM is either comparable to, or slightly better than ANN.

Decision tree based classifiers became extremely popular in machine learning after classification and regression trees (CART) were introduced by Breiman et al., (1984), although this type of classifier had been around since 1960's under the name of concept learning systems (CLS). In remote sensing applications, CART has been successfully used for the classification of multispectral and hyper-spectral images with high accuracy (e.g., Hansen et al., 1996; Friedl and Brodley 1997; Yang et al. 2003). Refinements over CART (e.g., bagging, boosting and random forests) have also been used in remote sensing for more accurate class label identification (e.g., Lawrence et al. 2004). In this paper, we illustrate that CART can sometimes yield unreliable predictions.

In this paper, we propose a new reliable multiclass-classifier for accurate class label prediction. The proposed classifier is based on the Bayesian *ensemble of trees* model called Bayesian Additive Regression Tree (BART) originally developed by Chipman et al. (2010). In the context of drug discovery and credit risk modeling, BART has been used for constructing binary classifiers under the name of BART probit (Chipman et al. 2010) and Bayesian Additive Classification Tree (BACT) (Zhang and Hardle 2010), respectively. In this paper, we follow a one-against-all approach for generalizing this binary classifier to a multiclass-classifier which is referred to as mBACT. As illustrated in the results section, mBACT yields more reliable predicted class labels than two popular competing classifiers, SVM and CART.

For performance comparison of different classifiers, we use a LANDSAT 5 TM image covering Kings County of Nova Scotia, Canada, acquired on August 15, 2009. This is a predominantly rural area with several small towns. The satellite (LANDSAT 5 TM) stopped working in November 2011 and the images over Kings County during September 2009 – November 2011 were unclear (i.e., cloudy or snowy). Thus, we used the scene acquired on August 15, 2009 (path/row: 08/29, L1T product, UTM map projection) for performance comparison. The satellite image consists of six reflectance bands (blue, green, red, near infrared and two middle infrared) with 30-meter resolution and one thermal band with 120-meter resolution. For maintaining resolution consistency, the thermal band was not used in building the classifier. Since the entire scene is very large ($185 \text{ km} \times 170 \text{ km}$), our study area consists of three relatively small regions in Kings County, i.e., the towns of Wolfville, Windsor and Kentville and their surrounding areas. Each image consists of six reflectance matrices (corresponding to the six bands), and each pixel can be classified in one of the seven land-use classes, built-up (BU), pond/lake/river water (W), Bay of Fundy (BF), agricultural land (AL), grassland (GL), trees (T) and scrubland (SL).

The remainder of the paper is organized as follows: The next section presents brief reviews on SVM, CART, BART and BART probit models. We propose the new classification methodology, mBACT, in the following section. Then, the data set obtained from LANDSAT 5 TM image and study areas are discussed. The following section starts with a brief review of the accuracy and uncertainty measures used for performance comparison. Then the classified images and tabulated results for the proposed and competing classifiers are presented. Finally, the paper concludes with the overall comparison of all classifiers, and a few important remarks.

## BACKGROUND

In this section we briefly review the key ideas of SVM and CART, and present the necessary details of BART methodologies that are relevant for the development of mBACT. For details on SVM, CART, BART and BACT see Liu and Zheng (2005), Brieman et al. (1984), Chipman et al. (2010) and Zhang and Hardle (2010) respectively.

### *Support Vector Machine (SVM)*

SVM was originally developed by Vapnik (1995) for binary classification, and was later extended to the domain of regression problems (Vapnik et al. 1996). A basic SVM classifier (designed for binary response) takes a set of inputs and predicts the class label for every given input. The main idea is to construct a separating hyperplane in the input space that can divide the training data into two classes with minimal error.

Let $Y$ be the response variable and $X = \{X^1, X^2, \ldots, X^p\}$ be the set of $p$ independent predictor variables. Suppose $D_0$ consists of $N$ training data points with binary response $y_i \in \{-1, 1\}$ at $p$-dimensional input $x_i = \left(x_i^1, x_i^2, \ldots, x_i^p\right)$, for $i = 1, \ldots, N$. Then, the linear separating hyperplane in the input space is $\{x : f(x) = w^T x + b = 0\}$, where $w \in R^p$, $b$ is a scalar and $f(x)$ is the decision function. That is, $\hat{f}(x) > 0$ implies that the predicted class label at $x$ is $\hat{y}(x) = 1$, whereas $\hat{f}(x) < 0$ supports $\hat{y}(x) = -1$. The parameters $(w, b)$ are obtained by solving the following optimization problem:

$$Minimize \quad L(w) = \frac{1}{2} ||w||^2 \qquad (1)$$

$$Subject\ to \quad y_i f(x_i) \geq 1, \quad i = 1, \ldots, N.$$

A linear hyperplane is often insufficient for separating the data into two classes, and a nonlinear hyperplane has to be constructed. This is achieved by transforming the input data in a much higher (possibly infinite) dimensional space called the *feature space* $(x \rightarrow \phi(x))$, and finding a linear hyperplane in that space. Typically, a kernel function $K(x_i, x_j)$ is used to implicitly define this non-linear transformation. Thus, the decision function becomes

$$f(x) = w^T \phi(x) + b = \sum_{i=1}^{N} \alpha_i y_i K(x_i, x) + b,$$

where $\alpha_i$ and $b$ are estimated by solving the optimization problem in Equation (1) with this decision function. A few popular kernels are as follows:

- Polynomial kernel: $K(x_i, x_j) = \left(\gamma \langle x_i, x_j \rangle + r\right)^d$

- Gaussian kernel: $K(x_i, x_j) = \exp\left(-\sigma |x_i - x_j|^2\right)$

- Sigmoid kernel: $K(x_i, x_j) = \tanh\left(\gamma \langle x_i, x_j \rangle + r\right)$

where $\gamma$ and $\sigma$ are scale parameters, $r$ is an offset parameter, and $d$ is the degree of the polynomial kernel.

In this paper, we are interested in a multiclass-classification problem with $n =$ seven classes (built-up, water, Bay of Fundy, agricultural land, grassland, trees and scrubland). The data set $D$ consists of $N$ points with response $y_i \in \{1, 2, \dots, n\}$ for every $p$-dimensional input $x_i = \left(x_i^1, x_i^2, \dots, x_i^p\right)$. One of the most popular technique for classifying multiclass data is to use the one-against-all approach (e.g., Bottou et al. 1994; Liu and Zheng 2005), where the main idea is to build a series of $n$ independent decision functions, $f_k(x)$, $k = 1, \dots, n$, and then choose the class label that corresponds to the largest $f_k$, i.e., $\hat{y}(x) = \arg \max \{f_k(x), \ k = 1, \dots, n\}$. Alternatively, one can follow the one-against-one approach (e.g., Knerr et al. 1990; Friedman 1996), in which all pairwise, $n(n-1)/2$, binary classifiers are trained and the class with maximum votes is the predicted class label $\hat{y}(x)$. Hsu and Lin (2002) argue that the one-against-one approach can often outperform the one-against-all method.

We used the built-in function $ksvm()$ in the R package "$kernlab$" (Karatzoglou et al. 2013) for fitting all SVM models. The classifier uses the one-against-one approach for solving a multiclass-classification problem. We used most of the default arguments in $ksvm()$ (including type="C-svc"), except the kernel parameters specified by kernel="polydot" and kpar = list(degree = 2, scale = 1, offset = 0.5). These values of the parameters in "kpar" have been chosen based on a preliminary study of the datasets considered in this paper. One may find alternative parameters combination to be more appropriate in other applications. As expected from any classifier, $ksvm()$ can produce both the predicted class label $\hat{y}(x)$ and the classification probability $\hat{p}_k(x) =$ $\hat{P}(Y = k|x)$ for every class $k = 1, \ldots, n$. These values are achieved by using type = "probabilities" and type = "response" in the built-in function $predict.svm()$.

It is worth noting that the predicted class label obtained via $ksvm()$ does not necessarily match with the maximum predicted classification probability $\hat{p}_{max}(x) = \max\{\hat{p}_k(x), k = 1, \ldots, n\}$, i.e., it is not always true that $\hat{y}(x) = \operatorname{argmax}\{\hat{p}_k(x), k = 1, \ldots, n\}$. It turns out that the two approaches, model prediction with type = "response", and model prediction via type = "probabilities", use different methods (see Wu et al. 2004 for details). For the examples considered in this paper, we implemented both approaches for predicting class labels, and realized that the first approach with type = "response" yields more accurate predicted class labels than the latter approach with predicted class label as $\hat{y}(x) = \operatorname{argmax}\{\hat{p}_k(x), k = 1, \ldots, n\}$. An illustration is given in the Results section.

### Classification and Regression Tree (CART)

Classification trees gained much popularity in the machine learning literature since CART was developed by Brieman et al. (1984). In the context of image classification, the main idea is to come up with a decision tree that partitions the image via recursive partitioning into homogeneous regions, for instance, built-up, water, trees, and so on. We only discuss binary trees, as the decision trees with multi-way splits can easily be obtained by iterative binary splits in binary trees.

The construction of a binary decision tree starts with assigning the entire training data $D$ ($N$ points) in one group called the root node. This node is then split into two nodes via one of the $p$ predictors. For instance, $X^i$ can be used to split the entire data into two subgroups or nodes $\{x : x^i \leq a\}$ and $\{x : x^i > a\}$. The two nodes are then further split using a value of another (or the same) predictor variable. The splitting process continues until a full tree is grown. Subsequently, techniques like cross-validation and tree complexity are used to prune the branches with very few data points to avoid over-fitting. Finally, each terminal (or leaf) node is assigned one class label $k \in \{1, 2, \ldots, n\}$.

Choosing the best splitting variable and split point combination $(X^i, a)$ at every node is an important part of the decision tree construction. An optimal $(X^i, a)$ combination is obtained by minimizing the total within partition miss-classification error $p_E(x) = 1 - p_{max}(x)$ (or some other impurity indices like Gini index or entropy) over every predictor-split point combination.

We used the implementation of CART in the R package "$rpart$" (Therneau et al. 2013) for all examples in this paper. First, $rpart()$ is used to grow the full tree, then $prune.rpart()$ prunes the tree using a tree complexity parameter ($cp$). Note that the value of $cp$ has to be provided in $rpart()$ as a control parameter, however, it can be tuned afterwards based on the cross validation error and the number of splits in the fully grown tree. Interestingly, for all examples considered in the paper, the optimal value of tree complexity parameter turns out to be $cp$ = 0.01 (the default value). We used mostly the default parameters of $rpart()$, and $prune.rpart()$, except minsplit = 10 (in $rpart.control$), which determines the minimum number of points a node must have to be considered for splitting, and xval = 5 (in $rpart.control$), which specifies 5-fold cross-validation. As in $ksvm()$, one can use $predict.rpart()$ with type = "class" and type = "prob" to obtain the predicted class labels $\hat{y}(x)$ and multiclass-classification probabilities $\{\hat{p}_k(x), k = 1, \dots, n\}$. Unlike $ksvm()$, here, $\hat{y}(x) = \text{argmax} \{\hat{p}_k(x), k = 1, \dots, n\}$ for all $x$ in the input space (image).

### Bayesian Additive Regression Tree (BART)

Chipman et al. (2010) proposed a flexible non-parametric regression model called BART, which is a sum of trees model developed in the Bayesian framework. This paper proposes a new multiclass-classifier (referred to as mBACT) based on BART models. In this section, we present a brief overview of BART.

The BART model represents the response $y$ at an input $x$ as a sum of $m$ adaptively chosen functions and an independent normal error,

$$y(x) = g(x; T_1, M_1) + g(x; T_2, M_2) + \cdots + g(x; T_m, M_m) + \varepsilon = h(x) + \varepsilon, \qquad (2)$$

where the function $g(x; T_j, M_j)$ denotes a "regression tree" model for a binary decision tree $T_j$ with $B_j$ terminal nodes (i.e., the input space is partitioned into $B_j$ rectangular regions), $\varepsilon \sim N(0, \sigma^2)$, and $M_j = \{\mu_{j1}, \mu_{j2}, \dots, \mu_{jB_j}\}$ is the collection of terminal node predictions for tree $T_j$. As a function of $x$, the $j$-th tree model $g(x; T_j, M_j)$ produces a piecewise-constant output, $\mu_{jb_j}$, which is obtained by following the sequence of decision rules in tree $T_j$ and arriving at the terminal node $1 \leq b_j \leq B_j$. The model prediction at any input $x$ is obtained by combining the ensemble of $m$ trees output, i.e., $h(x) = \sum_{j=1}^{m} \mu_{jb_j}$. Viewing Equation (2) as a statistical model, the parameter vector $\Theta = (T_1, \dots, T_m, M_1, \dots, M_m, \sigma)$ has to be estimated. Chipman et al. (2010) take large values of $m$ (50 to 200) and fit the model (2) in a Bayesian framework using Markov Chain Monte Carlo (MCMC) methods.

Each of the $m$ decision trees is constructed using an extension of the Bayesian CART methodology (Chipman et al. 1998), where the basic idea is to specify a prior on the tree space and a prior on the terminal nodes outputs for each tree in the tree space. Instead of a closed-form prior, a tree-generating stochastic process was used on the tree space. Combining this prior with the tree model likelihood yields a posterior distribution on the tree space. An efficient MCMC-based algorithm was used to stochastically search for good trees in the tree space. Since BART uses a sum of trees model, the trees with fewer splits (i.e., less than 5 splits) were assigned higher prior probabilities, and discrete uniform prior was used to choose the set of candidate split points for every split. For more details on these priors see Chipman et al. (2010).

We used the R library $BayesTree$ (Chipman and McCulloch 2009) for implementing BART (a key component of mBACT). The main function is $bart()$, which takes several arguments for controlling different features of the MCMC chains, and the priors on tree parameters, terminal node predictions, $\mu_{jb_j}$, and the noise variance. The arguments related to the MCMC chain specifies that the predictive samples are saved every "keepevery" rounds after "nskip" samples are discarded as burn-in, and the chain stops after "ndpost" realizations. A few important tree parameters are "numcut" − the maximum number of split points along each input, "ntree" – the number of trees in the ensemble ($m$), and "$k$" – the variance parameter in terminal node predictions, $\mu_{jb_j} \sim N(0, \sigma_\mu^2)$, where $\sigma_\mu \propto 1/k\sqrt{m}$. Prior on the noise variance ($\sigma^2$) can also be passed in via "sigest", "sigdf" and "sigquant".

As compared to a single tree based CART model, the BART model (2) is an "ensemble" of $m$ decision trees, and thus creates a flexible modeling framework. BART is also capable of incorporating higher dimensional interactions, by adaptively choosing the structure and individual rules of $T_j$'s. The sum of trees aspect of BART implicitly shares information from neighbouring inputs, and models the spatial dependency. Furthermore, many individual trees ($T_j$) may place split points in the same area, allowing the predicted function to change rapidly nearby, effectively capturing nonstationary (spiky) behaviour such as abrupt changes in the response (e.g., between roads, grassland and water).

### *BART as a Binary Classifier*

Like many other statistical regression models, BART can also be used for classification. Chipman et al. (2010) proposed an extension of BART called "*BART probit*" for classifying binary response $y \in \{0, 1\}$. The main idea is to model the response as

$$p(x) = P(Y = 1|x) = \Phi[h(x)],$$

where $h(x) = \sum_{j=1}^{m} g(x; T_j, M_j)$, and $\Phi$ is the standard normal cumulative distribution function (CDF), corresponding to a probit link function. Let $D_0$ be a training dataset with binary response $y_i \in \{0, 1\}$ at $p$-dimensional input $x_i = (x_i^1, x_i^2, \dots, x_i^p)$, for $i = 1, \dots, N$. For model convenience the response is rescaled to $[-3, 3]$. The BART probit model fit returns the posterior realizations of $T_j$'s and $M_j$'s, which are used to compute Monte Carlo estimate of $p(x)$. If the posterior draws from MCMC runs are $\left(T_j^{(s)}, M_j^{(s)}\right)$, $j = 1, \dots, m$; $s = 1, 2, \dots, S$, the trained classifier $\hat{y}(x)$ is given by

$$\hat{y}(x) = \begin{cases} 1, & if \ \frac{1}{S}\sum_{s=1}^{S} \Phi[h^{(s)}(x)] \geq 0.5 \ , \\ 0, & otherwise \end{cases} \tag{3}$$

where $h^{(s)}(x) = \sum_{j=1}^{m} g\left(x; T_j^{(s)}, M_j^{(s)}\right)$ and $\hat{p}(x) = \frac{1}{S}\sum_{s=1}^{S} \Phi\left[h^{(s)}(x)\right]$.

From an implementation viewpoint, the built-in function $bart()$ in the R library $BayesTree$ can also be used to fit BART probit models. It is however worth noting that for BART probit implementation a few arguments of $bart()$ are now either ignored by the function or have slightly different values. For instance, in case of binary response, the noise variance ($\sigma^2$) is fixed at 1, and as a result "sigest", "sigdf" and "sigquant" are ignored. The prior variance on the terminal node output is defined by $\sigma_\mu = 3/k\sqrt{m}$ as compared to $\sigma_\mu = 0.5/k\sqrt{m}$ (used in the regression framework). One can also include an argument called "binaryOffset" for offsetting the value of $h(x)$ away from zero.

Chipman et al. (2010) used BART probit model in a drug discovery application, where the goal was to classify compounds. Zhang and Hardle (2010) independently developed a similar adaptation of BART, applying it to the problem of classifying the solvency status of German firms based on their financial statement information. Zhang and Hardle called their classifier Bayesian Additive Classification Trees (BACT), and demonstrated that BACT outperforms CART and SVM in identifying insolvent firms.

## NEW METHODOLOGY: mBACT

In this section we propose a new reliable classifier called mBACT, which is a multiclass generalization of BART probit and BACT. The key idea is to use BART probit with the one-against–all approach for developing a multiclass-classifier. Though the motivating application considered in this paper comes from remote sensing literature, the classifier proposed here can be used in other applications as well.

Let $D$ be a set of $N$ points with response $y_i \in \{1, 2, \dots, n\}$ and $p$-dimensional input $x_i = \left(x_i^1, x_i^2, \dots, x_i^p\right)$, for $i = 1, \dots, N$. The method starts with transforming the data $D$ that would facilitate the one-against-all approach. For each class $k = 1, \dots, n$, we generate a pseudo data set $D_k$ with original $x_i$, and new response $y_{i(k)}$ defined as follows:

$$y_{i(k)} = \begin{cases} 1, & if\ y_i = k \\ 0, & if\ y_i \neq k. \end{cases} \tag{4}$$

In the spirit of Liu and Zheng (2005), we build $n$ binary classifiers using $D_1, D_2, \ldots, D_n$, and then combine them to obtain the desired multiclass-classifier. For $k = 1, 2, \ldots, n$, let $p_{(k)}(x) = P(Y_{(k)} = 1 | x)$ be the classification probability of input $x$ with class label 1 (indicator of class $k$ against others) under the binary data $D_k = \{(x_i, y_{i(k)}),\ i = 1, \ldots, N\}$ [note that $p_{(k)}(x)$ and $p_k(x)$ are slightly different quantities]. Then, the data set $D_k$ is used to fit the standard BART probit model

$$p_{(k)}(x) = P(Y_{(k)} = 1 | x) = \Phi[h_{(k)}(x)],$$

where $h_{(k)}(x) = \sum_{j=1}^m g_{(k)}(x; T_{j(k)}, M_{j(k)})$, and $\Phi$ is the standard normal CDF. The Monte Carlo estimate of $p_{(k)}(x)$ obtained from the posterior draws of $T_{j(k)}$'s and $M_{j(k)}$'s is

$$\hat{p}_{(k)}(x) = \frac{1}{S} \sum_{s=1}^S \Phi[h_{(k)}^{(s)}(x)].$$

The classification probabilities of these $n$ binary classifiers based on $D_1, D_2, \ldots, D_n$ can be combined to obtain the predicted class label under the original $n$-class dataset $D$,

$$\hat{y}(x) = \operatorname{argmax}\{\hat{p}_{(k)}(x), k = 1, \ldots, n\}. \tag{5}$$

Since $p_{(k)}(x)$ are binomial success probabilities for $n$ datasets and not multinomial probabilities for one dataset, the total $\sum_{k=1}^n \hat{p}_{(k)}(x)$ is not always 1. However, $\hat{p}_k(x)$ can be approximated using $\hat{p}_{(k)}(x)$ as follows

$$\hat{p}_k(x) \approx \frac{\hat{p}_{(k)}(x)}{\sum_{l=1}^n \hat{p}_{(l)}(x)}, \qquad for\ k = 1, \ldots, n. \tag{6}$$

From Equations (5) and (6), it is clear that the predicted class label matches with the maximum predicted classification probability $\hat{p}_{max}(x) = \max\{\hat{p}_k(x), k = 1, \ldots, n\}$.

The implementation of mBACT requires fitting BART probit models on $n$ pseudo data sets $D_1, D_2, \ldots, D_n$. For all examples considered in this paper, we used mostly the default arguments of $bart()$, except different MCMC parameters, keepevery = 20, ndpost = 5000 (for more stable posterior estimates), tree parameter, numcut = 1000 (for refined search of optimal split points) and variance parameter $k = 1$. The default value of $k = 2$ imposes considerable shrinkage to the individual terminal node output $\mu_{jb_j}$, whereas the proposed change ($k = 1$) increases the prior variance and applies less shrinkage (or smoothness) of the response. This is particularly important for our application as the land-use (e.g., the buildup area and water bodies) changes abruptly.

Next we discuss the satellite images used for the performance comparison of mBACT with two popular classifiers (SVM and CART) in remote sensing.

## STUDY AREA AND DATA COLLECTION

A multispectral satellite image acquired by LANDSAT 5 TM on August 15, 2009 (path/row: 08/29, L1T product, UTM map projection) over Nova Scotia, Canada, is considered for this comparative study. Although LANDSAT 5 TM data consists of seven bands, the sixth band is thermal with coarser resolution (120-meter) than the other six reflectance bands (30-meter), thus we used only the six reflectance bands (blue, green, red, near infrared and two middle infrared bands).

This is a large LANDSAT scene ($185\,\text{km} \times 170\,\text{km}$) covering (43.632 N, 63.266 W) to (45.579 N, 65.169 W), where each pixel is of $30m \times 30m$ resolution. Based on the consistency of land-use and accessibility of ground data, we chose three relatively small regions of Kings County (the towns of Wolfville, Windsor and Kentville and their surrounding rural areas) from this scene for performance comparison.

(i)     Wolfville area: A small portion of the scene covering (45.070 N, 64.334 W) to (45.098 N, 64.386 W), with $105 \times 134$ pixel image (Figure 1(b)).

(ii)     Windsor area: A medium size region of the scene covering (44.932 N, 64.102 W) to (44.995 N, 64.195 W) with $236 \times 239$ pixels (Figure 1(c)).

(iii)     Kentville area: A relative large region of the scene from (45.044 N, 64.418 W) to (47.117 N, 64.552 W) with $278 \times 349$ pixels (Figure 1(d)).

A False Color Composite (FCC) of the three study areas constructed using three bands (green, red and near infrared) are shown in Figure 1.

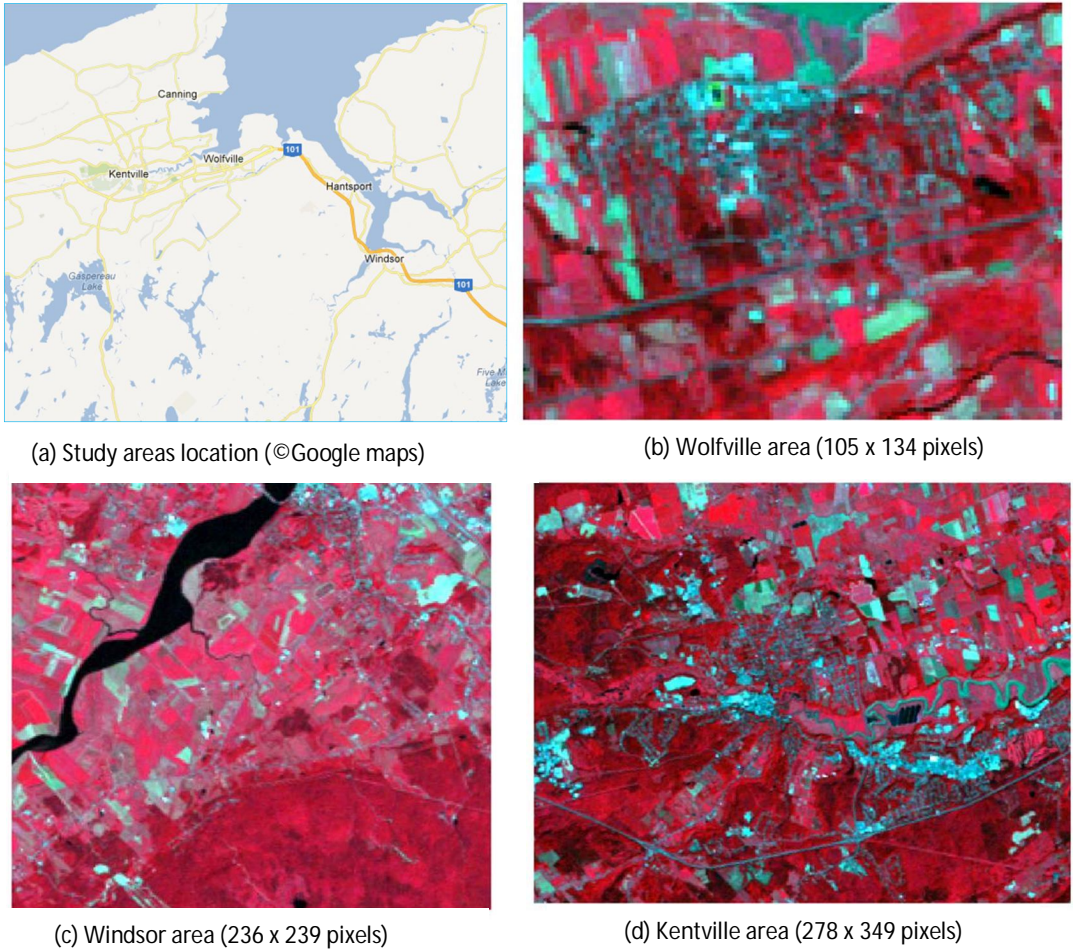

(a) Study areas location (©Google maps)

(b) Wolfville area (105 x 134 pixels)

(c) Windsor area (236 x 239 pixels)

(d) Kentville area (278 x 349 pixels)

Figure 1: A False Color Composite of the study areas in Nova Scotia.

Each pixel of the image in the three study areas can be classified into one of the seven classes, built-up, pond/lake/river water, Bay of Fundy, agricultural land or barren, grassland, trees and scrubland. The land-use of Windsor is interesting as it is clear from the map (Figure 1(a)) that the water body passing from the west to the north side of Figure 1(c) is a part of the Bay of Fundy that is cutoff due to the construction of a causeway for Highway 101. Since this water is not tidal, the "Bay of Fundy" class is not used in the Windsor scene.

The detailed class-by-class breakdown of the training and validation data size of the three study areas are shown in Table 1. The data were collected by sampling several disjoint homogeneous patches without replacement using class-wise stratified random sampling. Note that the sample size increases with the size of the region.

Table 1: Class-by-class distribution of the training and validation samples for each of seven classes in the three study areas.

| Wolfville area | | | |
|---|---|---|---|
| **Classes** | **Training** | **Validation** | **Total** |
| Built-up | 37 | 17 | 54 |
| Water | 12 | 10 | 22 |
| Bay of Fundy | 14 | 12 | 26 |
| Agricultural land | 31 | 14 | 45 |
| Grassland | 21 | 17 | 38 |
| Trees | 16 | 10 | 26 |
| Scrubland | 23 | 15 | 38 |
| **Total** | **154** | **95** | **249** |
| **Windsor area** | | | |
| Built-up | 76 | 42 | 118 |
| Water | 27 | 13 | 40 |
| Bay of Fundy | 0 | 0 | 0 |
| Agricultural land | 27 | 17 | 44 |
| Grassland | 61 | 31 | 92 |
| Trees | 61 | 32 | 93 |
| Scrubland | 69 | 33 | 102 |
| **Total** | **321** | **168** | **489** |
| **Kentville area** | | | |
| Built-up | 188 | 96 | 284 |
| Water | 35 | 19 | 54 |
| Bay of Fundy | 34 | 18 | 52 |
| Agricultural land | 78 | 46 | 124 |
| Grassland | 81 | 41 | 122 |
| Trees | 112 | 57 | 169 |
| Scrubland | 46 | 23 | 69 |
| **Total** | **574** | **300** | **874** |

For all three study areas, we followed the same approach of distributing the data points in the training and validation sets, that is, for each class, approximately two-thirds of the data points were assigned to the training set and the remainder to the validation set.

The data points (i.e., the ground truth or true class labels) were collected using a combination of in-person site visits, Google street views and Google satellite views. Although the data points were collected in 2012-2013, the land-use has not changed much (except a few differences in the built-up and scrubland classes) since 2009 (when the satellite image was taken). We want to emphasize that the main purpose of this paper is to "compare the classification accuracy" of mBACT with SVM and CART, which should not be affected by a few incorrectly labeled ground observations.

## RESULTS AND DISCUSSION

This section starts with a brief review of a few commonly used accuracy measures in the remote sensing literature for both the overall and class-wise comparison of the predicted class labels and ground truth at the validation data sites. Then we discuss a few measures based on $\hat{p}_k(x)$ for quantifying uncertainty in predicting the class labels. Finally, we present the classified images and tabulated results.

### *Accuracy Measures*

For a given classifier and a study area, let $F = \left( \left( f_{ij} \right) \right)$ be the error (or confusion) matrix, where $f_{ij}$ denotes the number of validation or reference points in the $j$-th class with predicted class labels $i$. Then, $f_{ii}$ is the number of correctly classified validation points in the $i$-th class, $f_{i+}$ ($i$-th row sum) is the number of validation points predicted to be in class $i$, and $f_{+i}$ ($i$-th column sum) is the true number of validation points in class $i$. Define $\eta_1 = (\sum_{i=1}^n f_{ii})/V$, $\eta_2 = (\sum_{i=1}^n f_{i+} f_{+i})/V^2$, $\eta_3 = [\sum_{i=1}^n f_{ii}(f_{i+} + f_{+i})]/V^2$, and $\eta_4 = [\sum_{i=1}^n \sum_{j=1}^n f_{ij}(f_{+i} + f_{j+})^2]/V^3$, where $V = \sum_{i=1}^n \sum_{j=1}^n f_{ij} = \sum_{i=1}^n f_{+i} = \sum_{i=1}^n f_{i+}$ is the grand total of the error matrix or the size of the validation data. Then, $\eta_1$ measures the *overall accuracy* of predicted class labels. Class-wise accuracies from user's and producer's perspectives can be measured by $f_{ii}/f_{i+}$ and $f_{ii}/f_{+i}$ respectively. The remaining quantities $\eta_2, \eta_3$ and $\eta_4$ are used in defining another popular accuracy measure called *kappa* ($\kappa$) which quantifies the agreement between the predicted class labels and the reality. The overall kappa coefficient is estimated by

$$\hat{\kappa} = \frac{\eta_1 - \eta_2}{1 - \eta_2},$$

and the associated uncertainty is measured by

$$var(\hat{\kappa}) = \frac{1}{V}\left\{\frac{\eta_1(1-\eta_1)}{(1-\eta_2)^2} + \frac{2(1-\eta_1)(2\eta_1\eta_2 - \eta_3)}{(1-\eta_2)^3} + \frac{(1-\eta_1)^2(\eta_4 - 4\eta_2^2)}{(1-\eta_2)^4}\right\}.$$

The prediction accuracy for $i$-th class can be measured by conditional kappa

$$\hat{\kappa}_i = \frac{V f_{ii} - f_{i+} f_{+i}}{V f_{i+} - f_{i+} f_{+i}},$$

with the associated variance

$$var(\hat{\kappa}_i) = \frac{V(f_{i+} - f_{ii})}{[f_{i+}(V - f_{+i})]^3}\{(f_{i+} - f_{ii})(f_{i+}f_{+i} - Vf_{ii}) + Vf_{ii}(V - f_{+i} - f_{i+} + f_{ii})\}.$$

Note that all of these accuracy measures are based on the discrepancy between the predicted and true class labels. Next, we present a few uncertainty measures that are based on the multiclass-classification probabilities.

### *Uncertainty Measures*

Assuming $\{\hat{p}_k(x), k = 1, \dots, n\}$ are estimated multinomial probabilities for the class labels of any pixel (or site) $x$ in the validation data, one can define the deviance as

$$D = -2\left(\sum_{i=1}^{V}\sum_{k=1}^{n} \log(\hat{p}_k(x_i)) \cdot I[\hat{y}(x_i) = k]\right),$$

where $I[\hat{y}(x_i) = k] = 1$, if the predicted class label is $k$, and zero otherwise. Small values of deviance represent confident prediction of correct class labels (i.e., less uncertainty). Since $\hat{y}(x)$ corresponds to $\max\{\hat{p}_k(x), k = 1, \dots, n\}$ for all $x$, in both CART and mBACT and not in SVM, it is expected that the deviances for CART and mBACT would be smaller as compared to that for SVM classified images.

We also use a few $p_k(x)$-based impurity indices for measuring uncertainty in predicting the class labels for every input site $x$ in the study area.

- Probability of miss-classification: $P_E(x) = 1 - \max\{p_k(x), \ k = 1, 2, \dots, n\}$
- Gini index: $G(x) = 1 - \sum_{k=1}^{n}[p_k(x)]^2$

- Entropy: $H(x) = - \sum_{k=1}^{n} p_k(x) \log \big( p_k(x) \big)$

In practice, we use the estimates of $p_k(x)$ for computing these impurity indices. Since none of these three measures are linked with the predicted class, we do not expect a systematic bias towards a particular classifier.

### *Classified Images and Tabulated Results*

We now compare the classified images obtained from SVM, CART and mBACT. Both overall and class-wise goodness-of-fit measures are tabulated and compared for these classifiers. Since the impurity indices ($P_E(x), G(x),$ and $H(x)$) are computed for every site in the study area, one can compare the uncertainty images instead of class-wise averages; however, due to limited space we only present uncertainty images for Kentville area. The results presented here are in the order of Wolfville, Windsor and Kentville (ordered based on the size of the region). At the end of this section, we will discuss the overall performance of the three classifiers.

Figure 2 shows the classified images of Wolfville area. It is clear from Figure 2 that both SVM and mBACT yield more accurate classified images as compared to CART. In particular, the predicted class labels for "built-up" and "trees" in CART appear to be relatively noisier as compared to that obtained from SVM and mBACT. A quick comparison of SVM with mBACT generated images does not show much difference. However, a closer look at the accuracy measures (Table 2) reveals more precise information and a clear overall trend in the prediction accuracy and uncertainty.

It is clear from Table 2 that in terms of overall *accuracy measures* ($\eta_1$ and $\kappa$), SVM generates the most accurate predicted class label in the validation set. Furthermore, mBACT is better than CART but inferior to SVM (i.e., SVM > mBACT > CART). Class-wise accuracy measures (user's and producer's accuracy and condition kappa) do not exhibit a consistent trend over the seven classes.

In terms of overall *uncertainty measures* (deviance, Gini and entropy)*,* Table 2 shows a different overall trend (CART > mBACT > SVM). Unlike accuracy measures, class-wise uncertainty measures (Gini and entropy) show the same consistent trend (CART > mBACT > SVM) for every class as well. By combining the information on accuracy and uncertainty measures, it appears that CART is a more confident but less accurate predictor than mBACT (mBACT > CART in terms of $\eta_1$ and $\kappa$). Since the predicted class labels do not necessarily match with the maximum predicted classification probabilities, it is not surprising that SVM leads to the largest deviance. However, the ranking of SVM based on overall and class-wise Gini and entropy values indicate that SVM-based predictions are more uncertain than mBACT.

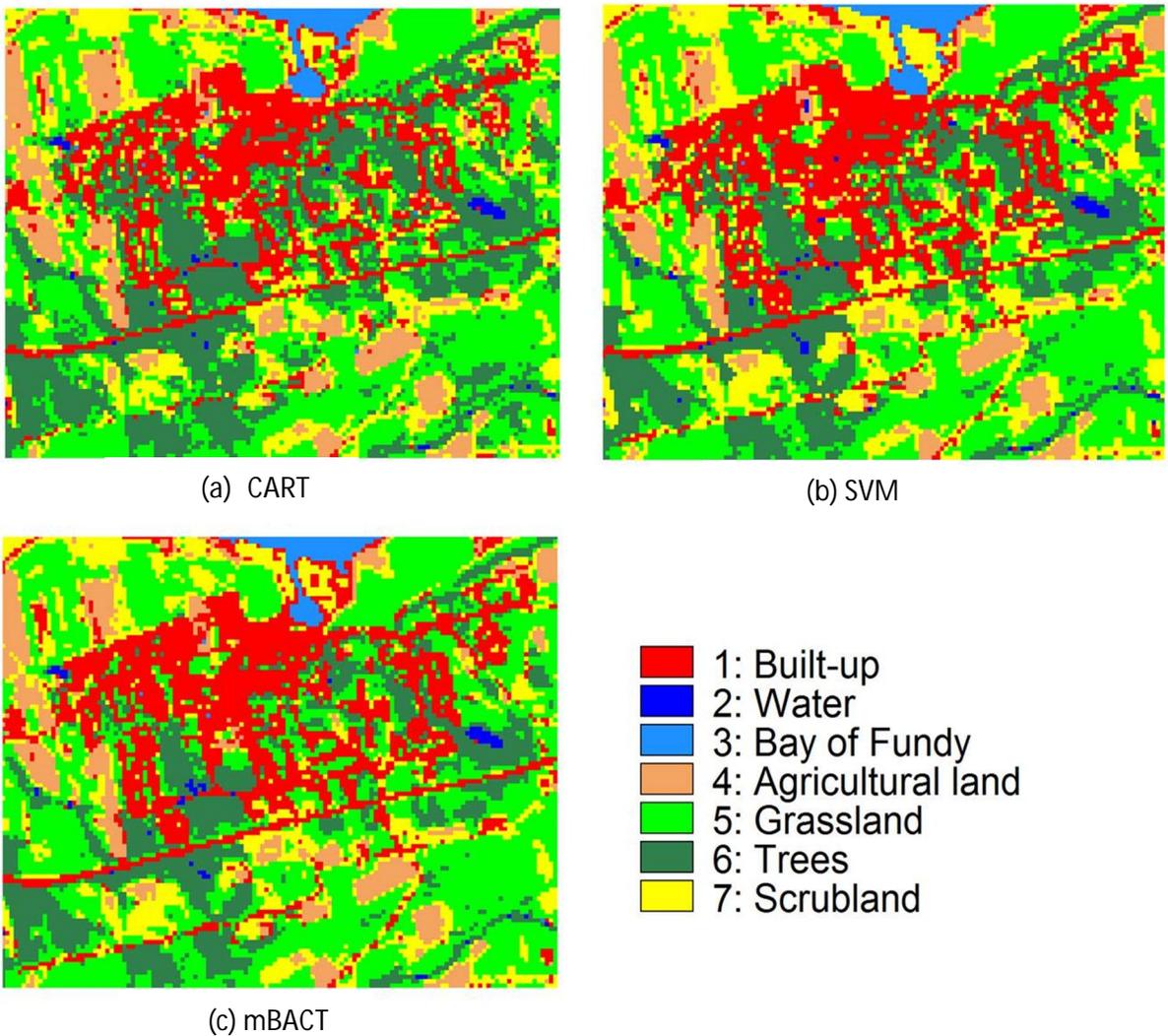

(a) CART

(b) SVM

1: Built-up
2: Water
3: Bay of Fundy
4: Agricultural land
5: Grassland
6: Trees
7: Scrubland

(c) mBACT

Figure 2: Classified images of Wolfville area

Table 2: Overall and class-wise accuracy and uncertainty measures for Wolfville.

| CART-Wolfville | | | | | |
|---|---|---|---|---|---|
| Overall $\eta_1$=83.16%  ,  Kappa ($\kappa$)=0.802  ,   Var(Kappa)=0.0020<br>Overall Gini = 0.127 ,  Overall Entropy  = 0.236,  Overall deviance = 12.03 | | | | | |
| Classes | User's | Producer's | Co. Kappa | Gini | Entropy |
| Built-up | 80.95 | 100.00 | 0.768 | 0.000 | 0.000 |
| Water | 100.00 | 80.00 | 1.000 | 0.000 | 0.000 |
| Bay of Fundy | 100.00 | 75.00 | 1.000 | 0.000 | 0.000 |
| Agricultural land | 59.09 | 92.86 | 0.520 | 0.233 | 0.462 |
| Grassland | 100.00 | 94.12 | 1.000 | 0.135 | 0.211 |
| Trees | 76.92 | 100.00 | 0.742 | 0.111 | 0.224 |
| Scrubland | 100.00 | 40.00 | 1.000 | 0.234 | 0.470 |
| SVM-Wolfville | | | | | |
| Overall $\eta_1$= 88.42%  ,   Kappa ($\kappa$)=0.864   ,   Var(Kappa)=0.0015<br>Overall Gini = 0.599 ,  Overall Entropy  = 1.310, Overall deviance = 110.69 | | | | | |
| Classes | User's | Producer's | Co. Kappa | Gini | Entropy |
| Built-up | 94.12 | 94.12 | 0.928 | 0.535 | 1.172 |
| Water | 90.00 | 90.00 | 0.888 | 0.531 | 1.147 |
| Bay of Fundy | 100.00 | 91.67 | 1.000 | 0.685 | 1.485 |
| Agricultural land | 66.67 | 100.00 | 0.609 | 0.763 | 1.653 |
| Grassland | 94.12 | 94.12 | 0.928 | 0.516 | 1.131 |
| Trees | 100.00 | 90.00 | 1.000 | 0.626 | 1.375 |
| Scrubland | 90.00 | 60.00 | 0.881 | 0.771 | 1.684 |
| mBACT-Wolfville | | | | | |
| Overall $\eta_1$=84.21%  ,    Kappa ($\kappa$)=0.814 ,  Var(Kappa)=0.0019<br>Overall Gini = 0.471 ,  Overall Entropy  = 0.856, Overall deviance = 48.62 | | | | | |
| Classes | User's | Producer's | Co. Kappa | Gini | Entropy |
| Built-up | 80.00 | 94.12 | 0.756 | 0.558 | 0.917 |
| Water | 100.00 | 80.00 | 1.000 | 0.501 | 0.933 |
| Bay of Fundy | 100.00 | 83.33 | 1.000 | 0.282 | 0.758 |
| Agricultural land | 66.67 | 100.00 | 0.609 | 0.372 | 0.718 |
| Grassland | 89.47 | 100.00 | 0.872 | 0.385 | 0.792 |
| Trees | 90.91 | 100.00 | 0.898 | 0.536 | 0.901 |
| Scrubland | 83.33 | 33.33 | 0.802 | 0.566 | 0.963 |

The classified images of Windsor area are shown in Figure 3. A quick view of Figure 3 shows that all three classifiers perform reasonably well in capturing the main features of the land-use. See Table 3 for a detailed performance comparison of the classifiers.

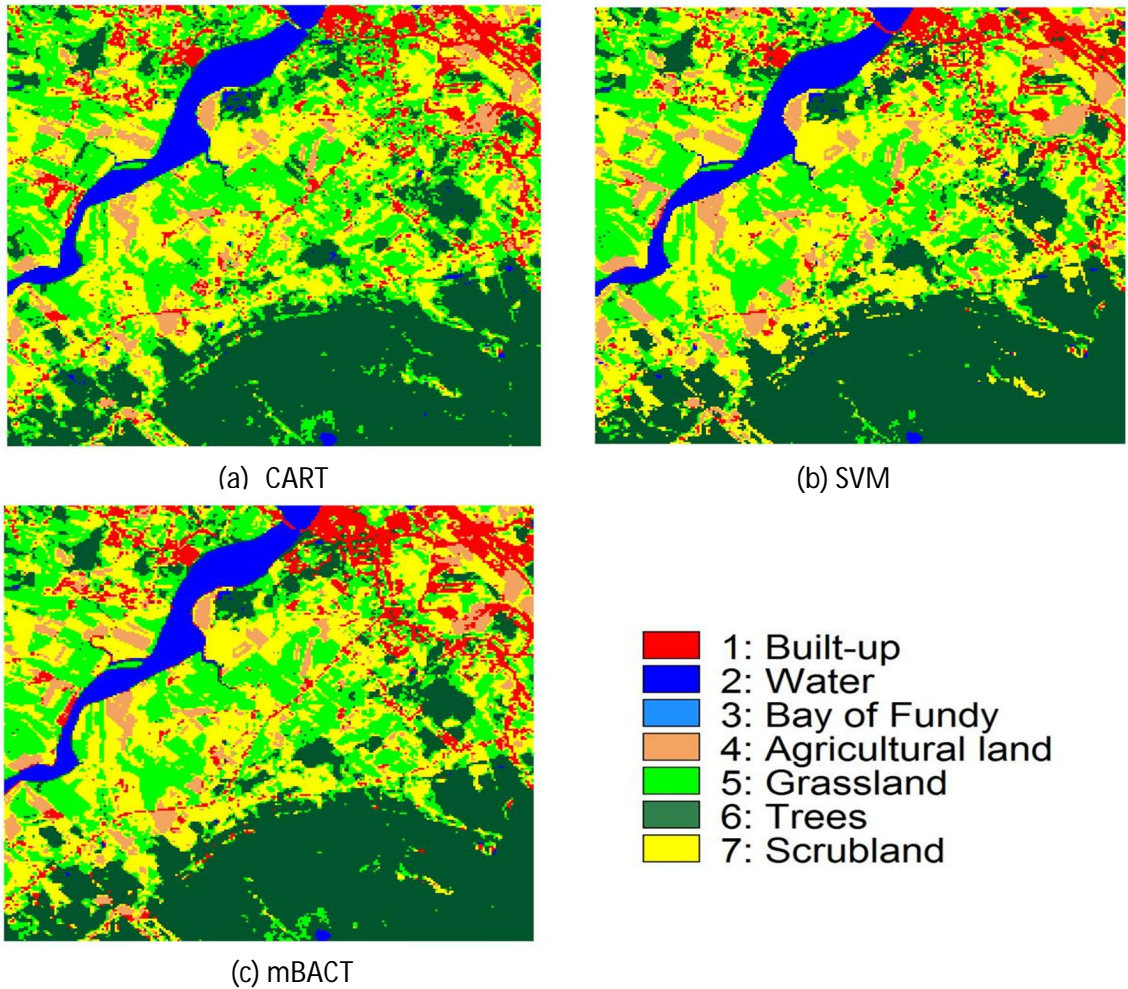

(a) CART

(b) SVM

(c) mBACT

1: Built-up
2: Water
3: Bay of Fundy
4: Agricultural land
5: Grassland
6: Trees
7: Scrubland

Figure 3: Classified images of Windsor area

In terms of both overall accuracy measures ($\eta_1$ and $\kappa$), it is clear from Table 3 that mBACT is slightly better than SVM and much better than CART. In this case, even the class-wise accuracy measures (user's, producer's and conditional kappa) support the superior performance of mBACT in most of the classes. The uncertainty measures (deviance, Gini and entropy) follow the same trend as in the Wolfville image.

Table 3: Overall and class-wise measures of accuracy and uncertainty for Windsor.

| CART-Windsor | | | | | |
|---|---|---|---|---|---|
| Overall $\eta_1$= 87.50%  ,    Kappa ($\kappa$)= 0.847  ,    Var(Kappa)=0.0010<br>Overall Gini  = 0.130  ,   Overall Entropy  =  0.268, Overall deviance = 22.99 | | | | | |
| Classes | User's | Producer's | Co. Kappa | Gini | Entropy |
| Built-up | 97.06 | 78.57 | 0.961 | 0.182 | 0.363 |
| Water | 86.67 | 100.00 | 0.856 | 0.069 | 0.154 |
| Bay of Fundy | - | - | - | - | - |
| Agricultural land | 87.50 | 82.35 | 0.861 | 0.180 | 0.325 |
| Grassland | 80.00 | 90.32 | 0.755 | 0.191 | 0.335 |
| Trees | 93.10 | 84.38 | 0.915 | 0.062 | 0.143 |
| Scrubland | 82.05 | 96.97 | 0.775 | 0.134 | 0.329 |
| **SVM- Windsor** | | | | | |
| Overall $\eta_1$=92.26%    ,   Kappa ($\kappa$)=0.905  ,    Var(Kappa)=0.0006<br>Overall Gini = 0.463 ,  Overall Entropy  = 0.990, Overall deviance = 129.30 | | | | | |
| Classes | User's | Producer's | Co. Kappa | Gini | Entropy |
| Built-up | 95.00 | 90.48 | 0.933 | 0.446 | 0.903 |
| Water | 100.00 | 100.00 | 1.000 | 0.117 | 0.254 |
| Bay of Fundy | - | - | - | - | - |
| Agricultural land | 100.00 | 76.47 | 1.000 | 0.594 | 1.249 |
| Grassland | 90.62 | 93.55 | 0.885 | 0.491 | 1.058 |
| Trees | 87.88 | 90.62 | 0.850 | 0.350 | 0.773 |
| Scrubland | 89.19 | 100.00 | 0.866 | 0.592 | 1.250 |
| **mBACT- Windsor** | | | | | |
| Overall $\eta_1$=93.45%        ,    Kappa ($\kappa$)=0.919  ,   Var(Kappa)=0.0005<br>Overall Gini = 0.311  ,  Overall Entropy  = 0.492, Overall deviance = 49.73 | | | | | |
| Classes | User's | Producer's | Co. Kappa | Gini | Entropy |
| Built-up | 95.00 | 90.48 | 0.933 | 0.413 | 0.603 |
| Water | 100.00 | 100.00 | 1.000 | 0.107 | 0.385 |
| Bay of Fundy | - | - | - | - | - |
| Agricultural land | 93.75 | 88.24 | 0.931 | 0.364 | 0.648 |
| Grassland | 93.33 | 90.32 | 0.918 | 0.362 | 0.534 |
| Trees | 93.75 | 93.75 | 0.923 | 0.201 | 0.333 |
| Scrubland | 89.19 | 100.00 | 0.866 | 0.389 | 0.598 |

Figure 4 displays the classified images of Kentville area. It is clear from Figure 4 that CART yields somewhat noisier classified map (particularly in built-up and scrubland classes) than SVM and mBACT.

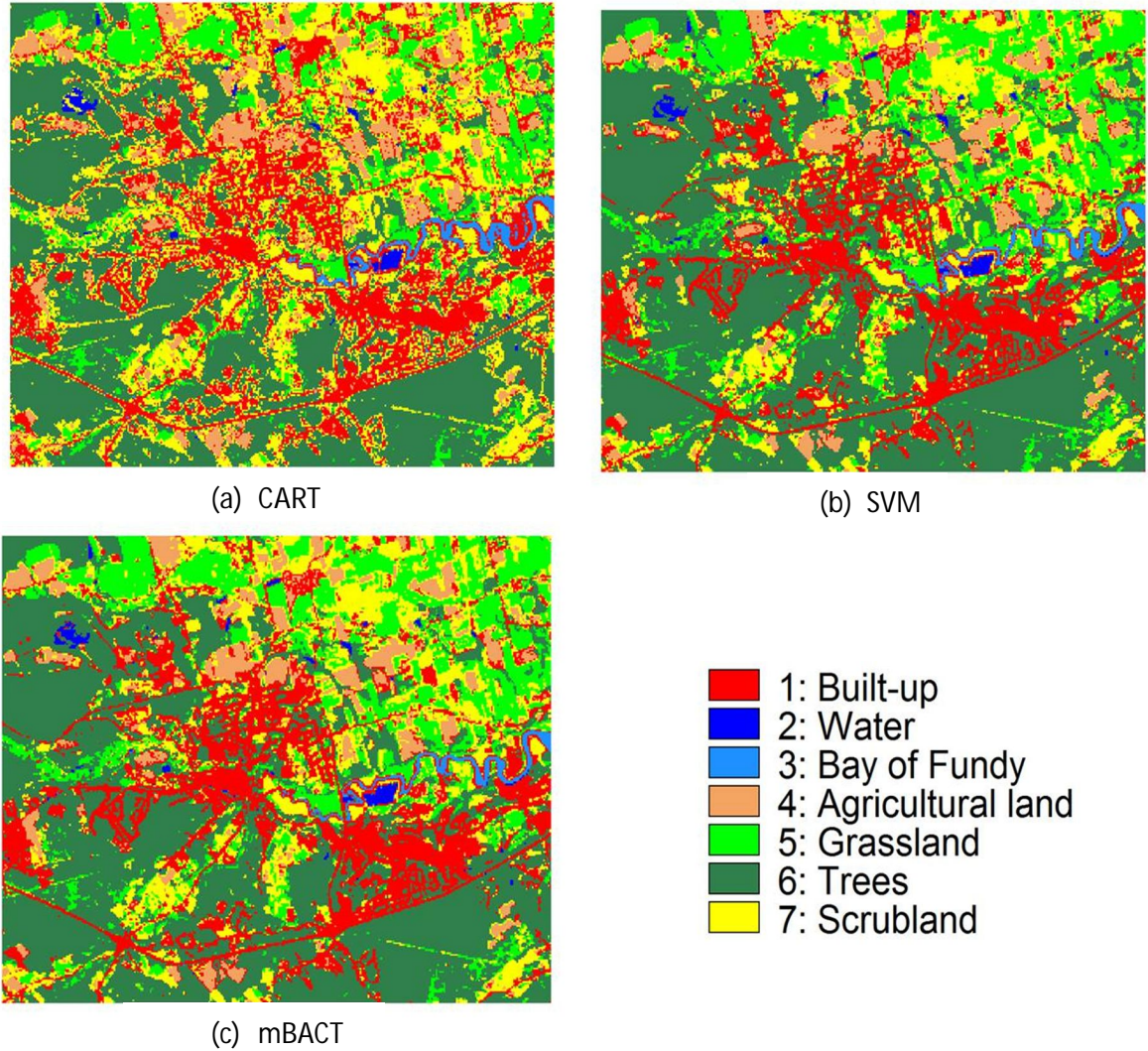

(a) CART

(b) SVM

(c) mBACT

■ 1: Built-up
■ 2: Water
■ 3: Bay of Fundy
■ 4: Agricultural land
■ 5: Grassland
■ 6: Trees
■ 7: Scrubland

Figure 4: Classified images of Kentville area

Goodness-of-fit measures for Kentville area images are summarized in Table 4. The overall accuracy ($\eta_1$ and $\kappa$) indicate that mBACT is comparable to SVM and produces more accurate predicted class labels than CART. Similar to the Wolfville image, the class-wise accuracy measures (user's, producer's and conditional kappa) do not exhibit a clear trend across the classifiers, but the uncertainty measures show a consistent pattern with CART being the most confident classifier and SVM the most uncertain.

Note that the value of deviance has been increasing with the size of the validation set, because it is a sum and not an average.

Table 4: Overall and class-wise measures of accuracy and uncertainty for Kentville.

| CART-Kentville | | | | | |
|---|---|---|---|---|---|
| Overall $\eta_1$= 84.33% ,    Kappa ($\kappa$)=0.807,    Var(Kappa)=0.0007<br>Overall Gini= 0.204,   Overall Entropy= 0.418, Overall deviance =50.29 | | | | | |
| Classes | User's | Producer's | Co. Kappa | Gini | Entropy |
| Built-up | 83.33 | 83.33 | 0.755 | 0.126 | 0.281 |
| Water | 95.00 | 100.00 | 0.947 | 0.059 | 0.136 |
| Bay of Fundy | 82.35 | 77.78 | 0.812 | 0.171 | 0.405 |
| Agricultural land | 84.85 | 60.87 | 0.821 | 0.353 | 0.665 |
| Grassland | 90.24 | 90.24 | 0.887 | 0.106 | 0.211 |
| Trees | 100.00 | 86.49 | 1.000 | 0.036 | 0.103 |
| Scrubland | 52.63 | 86.96 | 0.487 | 0.517 | 1.011 |
| SVM- Kentville | | | | | |
| Overall $\eta_1$= 90.00% ,    Kappa ($\kappa$)=0.875 ,    Var(Kappa)=0.0005<br>Overall Gini= 0.716,   Overall Entropy = 1.571, Overall deviance = 393.15 | | | | | |
| Classes | User's | Producer's | Co. Kappa | Gini | Entropy |
| Built-up | 89.11 | 93.75 | 0.840 | 0.507 | 1.147 |
| Water | 89.47 | 89.47 | 0.888 | 0.444 | 0.948 |
| Bay of Fundy | 100.00 | 83.33 | 1.000 | 0.721 | 1.577 |
| Agricultural land | 92.31 | 78.26 | 0.909 | 0.778 | 1.726 |
| Grassland | 88.10 | 90.24 | 0.862 | 0.725 | 1.551 |
| Trees | 91.80 | 98.25 | 0.899 | 0.783 | 1.723 |
| Scrubland | 82.61 | 82.61 | 0.812 | 0.776 | 1.676 |
| mBACT- Kentville | | | | | |
| Overall $\eta_1$=90.00% ,    Kappa ($\kappa$) =0.875 ,    Var(Kappa)=0.0006<br>Overall Gini = 0.365 ,  Overall Entropy = 0.619, Overall deviance = 106.67 | | | | | |
| Classes | User's | Producer's | Co. Kappa | Gini | Entropy |
| Built-up | 87.38 | 93.75 | 0.814 | 0.421 | 0.671 |
| Water | 86.36 | 100.00 | 0.854 | 0.359 | 0.601 |
| Bay of Fundy | 100.00 | 72.22 | 1.000 | 0.258 | 0.619 |
| Agricultural land | 90.00 | 78.26 | 0.882 | 0.435 | 0.715 |
| Grassland | 88.37 | 92.68 | 0.865 | 0.384 | 0.649 |
| Trees | 100.00 | 96.49 | 1.000 | 0.245 | 0.460 |
| Scrubland | 79.17 | 82.61 | 0.774 | 0.564 | 0.902 |

Figure 5 presents site-wise comparison of the probability of miss-classification $p_E(x)$, Gini index $G(x)$, and entropy $H(x)$ for all three classifiers. All uncertainty images support the expected trend, that is, CART is the most confident classifier, and SVM based predictions are most uncertain.

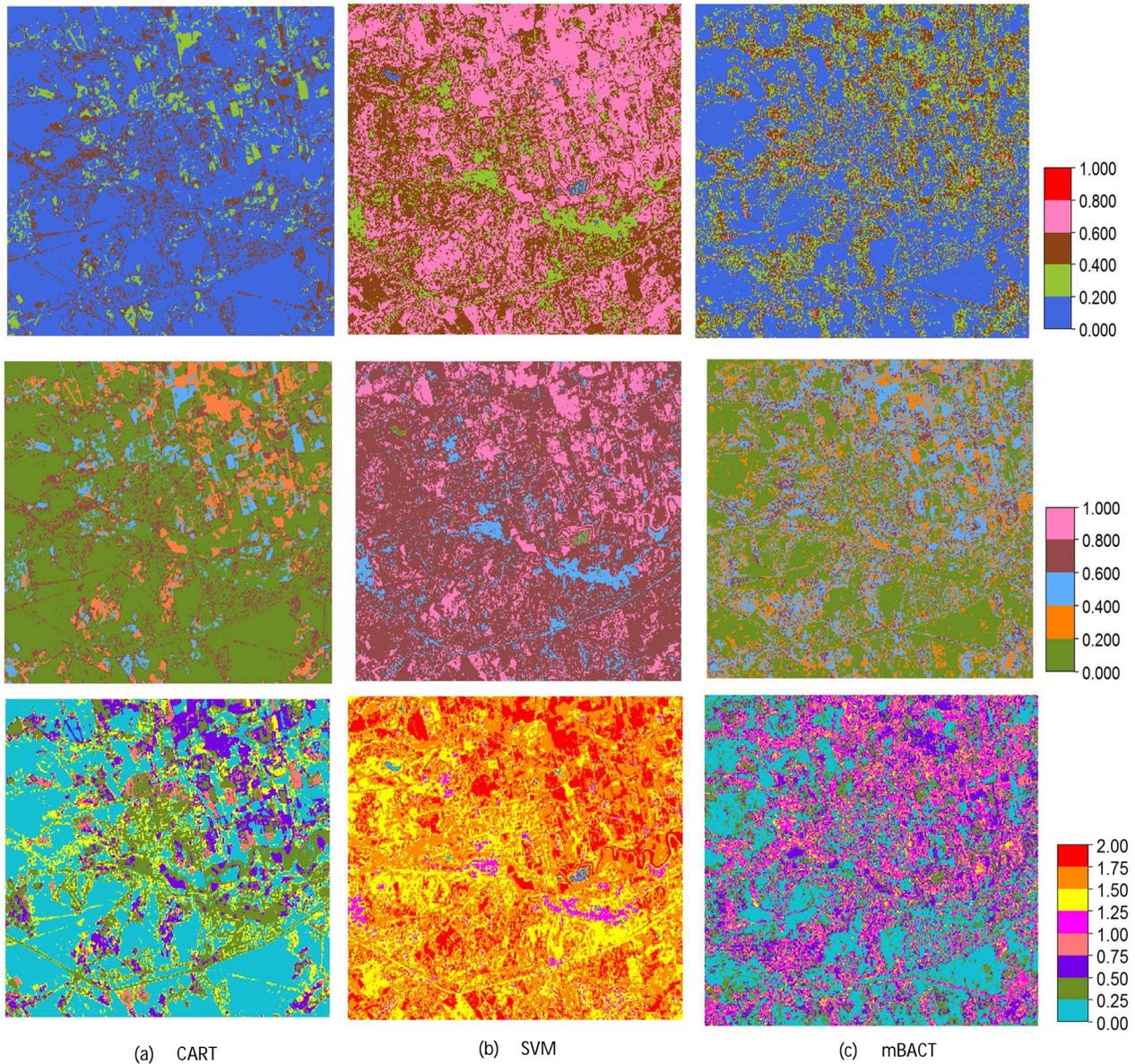

Figure 5: Uncertainty images of Kentville area. The first row of plots displays probability of miss-classification; the second row shows the Gini index plots; and the third row depicts the entropy.

***Overall Summary***

The performance comparison of mBACT with SVM and CART based on the three study areas suggest the following:

1. Based on the overall accuracy measures ($\eta_1$ and $\kappa$), mBACT outperforms CART in all cases, and performs same or better than SVM in two out of three cases. In one case (Wolfville), SVM yielded higher overall accuracy than mBACT.

2. In terms of uncertainty measures (deviance, Gini and entropy), mBACT generates predictions with slightly larger $\hat{p}_{max}(x)$ as compared to SVM, and CART turns out to be the most confident predictor.

Given that CART yields the smallest values of $\eta_1$ and $\kappa$ for all three images, the over confidence in terms of deviance, Gini and entropy is somewhat questionable. To investigate this further we present a display which evaluates the accuracy of prediction of the class probabilities. This is accomplished by comparing the maximum predicted probability $\hat{p}_{max}(x)$ with the actual class label Y for each observation in the validation set. A direct comparison between $\hat{p}_{max}(x)$ and Y at the level of individual observations is not practical, since the observed class label will either equal the class that has maximum predicted probability, or not. In other words, such a comparison would be between a predicted probability and a binary indicator for whether the observed class is the same as the predicted class. However, such a comparison can be made by combining observations into groups, and comparing the proportion of correctly predicted classes to the average value of $\hat{p}_{max}(x)$ in each group. We group the observations as follows: For each observation, obtain $\hat{p}_{max}(x)$. Then sort the observations from smallest to largest value of $\hat{p}_{max}(x)$. Divide these sorted observations into 10 groups. Thus the 1[st] group will consist of observations with the smallest amount of certainty in prediction, and the 10[th] group will consist of observations with the largest prediction certainty. For each group, also calculate the proportion of observations for which the class with maximum predicted probability was equal to the observed class. If a model is accurately predicting the class probabilities, then this proportion of correctly classified observations in each group should equal the mean of the maximum predicted class probabilities. A plot of these quantities for the 10 groups should correspond to a line with intercept 0 and slope 1, for the ideal predictor. Figure 6 presents a comparison of such plot for all three classifiers in the three study areas.

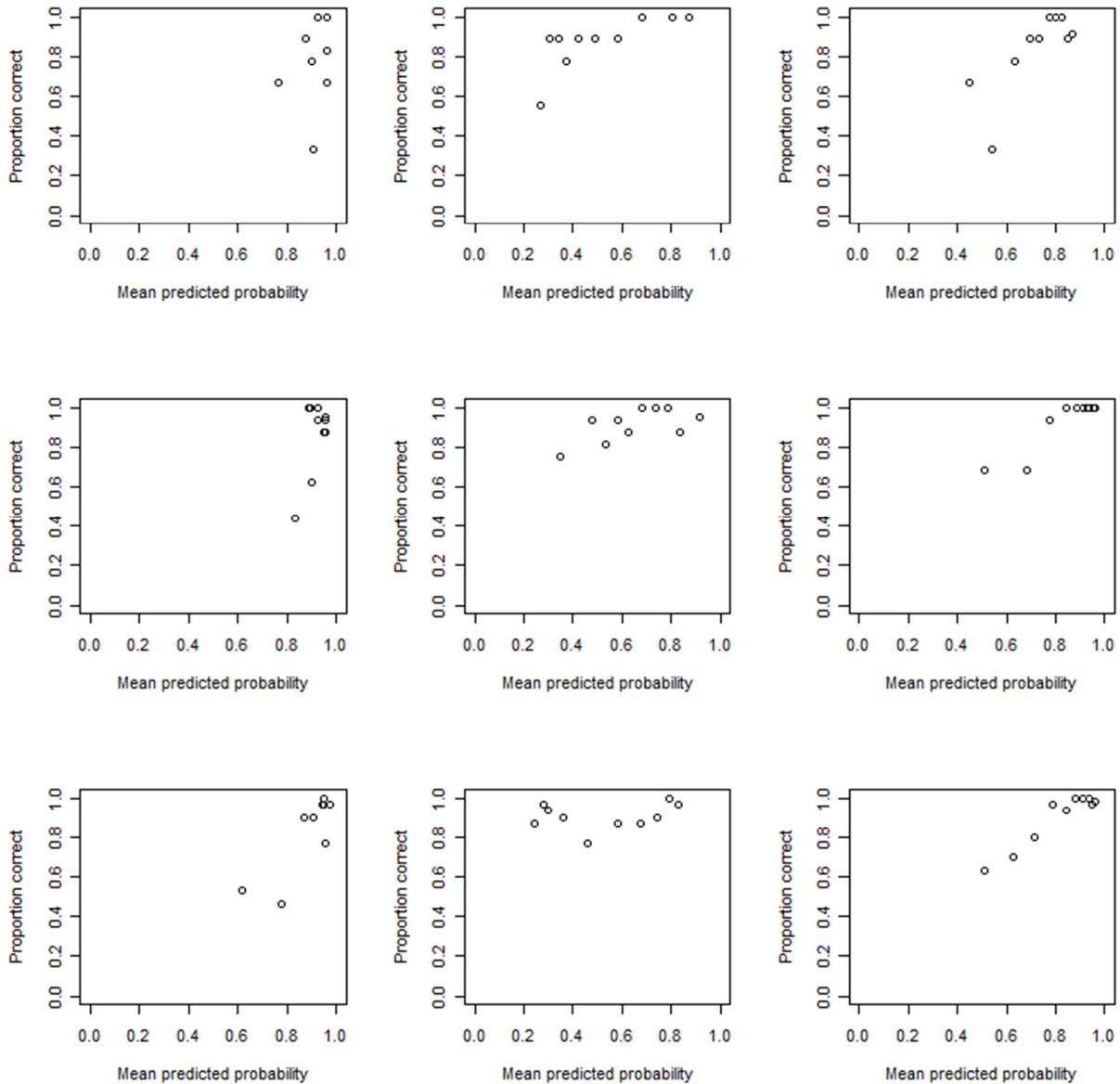

Figure 6: [1st row: Wolfville, 2nd row: Windsor, 3rd row Kentville; 1st column: CART, 2nd column SVM, 3rd column mBACT]. Each plot presents "proportion of correctly classified validation points" versus "average $\hat{p}_{max}(x)$" in 10 $\hat{p}_{max}(x)$ quantile bins over the validation set.

It is clear from Figure 6 that mBACT (the third column) is the most reliable classifier in this measure (with a slope closest to 1), and CART (the first column) leads to most confident prediction (with points whose mean predicted $\hat{p}_{max}(x)$ exceed the proportion correctly classified within each group). Note that in mBACT-Wolfville plot, the smallest "proportion of correct classification" point with 30% correct classification corresponds to average $\hat{p}_{max}(x) \approx 0.55$, whereas in CART-Wolfville, average $\hat{p}_{max}(x) \approx 0.9$. As a result CART can be an unreliable (i.e. overconfident) classifier.

Although we include plots in Figure 6 for SVM, we note that the SVM package uses two different models to make a class prediction and to predict class probabilities. Thus there is less reason to expect that the points in the middle column of Figure 6 will have a line with slope 1, since the class labels used in calculating the proportion of correct classifications (vertical axis) are from a different model than the predicted class probabilities. Indeed, there seems to be a very weak correspondence between $\hat{p}_{max}(x)$ and the proportion of correct classifications for SVM in Figure 6.

## CONCLUDING REMARKS

Accurate prediction of class labels of a satellite image has been a challenging problem in remote sensing applications. In this article, we introduced a new reliable multiclass-classifier, mBACT, for accurate identification of class labels. Though the classification problem considered in this paper comes from a remote sensing application, mBACT can be used for other applications as well.

The main idea of mBACT was to generalize the binary classifier (BART probit or BACT) for multiclass-classification problem using the one-against-all approach. This requires fitting BART model $n$ times (the number of classes) for the entire data set, and the current version of the $R$ library $BayesTree$ (for fitting BART models) is computationally more expensive than SVM and CART. This is somewhat expected because BART is a Bayesian ensemble of trees model, and CART is based on a single decision tree. The authors of $BayesTree$ library are currently working on using parallel computation tools to speed up the implementation process. Subsequently, it might be worth investigating the relative performance of mBACT with one-against-one generalization of BART probit.

## ACKNOWLEDGMENT

We would like to thank the referees for many useful comments and suggestions that led to significant improvement of the article. This work was supported in part by Discovery grants from the Natural Sciences and Engineering Research Council of Canada.